\documentstyle[preprint,prl,aps]{revtex}
\tightenlines
\begin{document}
\preprint{\it submitted to Physical Review Letters}
\title{Spin diffusion in doped
semiconductors}
\author{Michael E. Flatt\'e}
\address{Department of Physics and Astronomy, University of
Iowa, Iowa City, Iowa 52242}
\author{Jeff M. Byers}
\address{Naval Research Laboratory, Washington, D.C.  20375}
\maketitle
\begin{abstract}
The behavior of spin diffusion in doped semiconductors is
shown to be qualitatively different than in undoped
(intrinsic) ones. Whereas a spin packet in an intrinsic
semiconductor must be a multiple-band disturbance, involving
inhomogeneous distributions of both electrons and holes, in
a doped semiconductor a single-band disturbance is possible.
For $n$-doped nonmagnetic semiconductors the enhancement of
diffusion due to a degenerate electron sea in the conduction
band is much larger for these single-band spin packets than
for charge packets,
and can exceed an order of magnitude at low
temperatures
even for equilibrium
dopings as small as 10$^{16}$~cm$^{-3}$. In $n$-doped 
ferromagnetic and semimagnetic semiconductors the motion of spin packets
polarized antiparallel to the equilibrium carrier spin
polarization is
predicted to be an order of magnitude faster than for
parallel polarized spin packets. These results are
reversed for $p$-doped semiconductors.
\end{abstract}
\vfill\eject

The motion and persistence of inhomogeneous electronic
distributions are central to the
electronic technologies based on semiconductors. Recently a
broader category of possible disturbances, namely those
involving inhomogeneous spin distributions in doped semiconductors, 
have been shown
to exhibit long lifetimes\cite{Wagner,Aws_PRL,Aws_nature} and 
anomalously high diffusion
rates\cite{Aws_nature}. This behavior
indicates the potential of a new
electronic technology\cite{Datta} relying 
on spin. A crucial requirement of this new 
technology, however, is the clarification of the transport
properties of inhomogeneous spin distributions. \cite{Aws_PT} 
A full understanding is also desirable of the relationship
between the physical effects driving
semiconductor spin electronics and those driving the mature 
area of metallic spin
electronics,\cite{Prinz} which has produced advances in
magnetic read heads and non-volatile memory.

These spin distributions are also of fundamental interest as
well, for they are phase-coherent states which 
can be very long lived ($>100$~ns) and very
extended ($>100\ \mu$m). In contrast to
phase-coherent ground states, such as the BCS ground state of a
superconductor or the Laughlin state of the fractional
quantum hall system, these spin distributions are
nonequilibrium phase-coherent states. Their long lifetime and large spatial
size allow unprecedented probes of phase-coherent
behavior --- of which Refs.~\cite{Aws_PRL,Aws_nature} are
initial examples.

We consider the properties of doped and undoped
semiconductors which are unpolarized in equilibrium but
have a localized perturbation of the carriers. In the
highly-doped limit this system should behave like a
paramagnetic metal (such as the copper used in Co/Cu multilayer
giant magnetoresistive devices\cite{Parkin}).
Qualitative differences in diffusion are found between the
doped and undoped systems, at doping densities of
10$^{16}$~cm$^{-3}$ at low temperature and
10$^{18}$~cm$^{-3}$ at room temperature. Quantitative
agreement is found with recent experimental results on rapid
spin diffusion at low temperature. \cite{Aws_nature} We also describe
spin diffusion in spin polarized semiconductors.
This work may assist in understanding 
spin transport within metallic
ferromagnetic semiconductors, such as
\hbox{GaMnAs},\cite{GaMnAs} which has been used in 
spin-dependent resonant tunneling devices,\cite{Ohno} and
semimagnetic semiconductors, such as 
BeMnZnSe,\cite{BeMnZnSe} which has been used in a
spin-polarized light-emitting diode.\cite{spinLED}

The origin of the differences in spin diffusion
between semiconductors and metals are (1) the much greater
spin relaxation lifetime in semiconductors, (2) the relative
ineffectiveness of screening in semiconductors relative to
metals, and (3) the possibility of controlling whether
carriers in a band are degenerate or not by small perturbations
(e.g. electric fields or doping).
The first of these differences was explored in
Ref.~\cite{Aws_PT}. In this letter we examine the
implications of the second and third aspects. We show
that careful consideration of the consequences of (2) and
(3) lead to a direct explanation of the anomalously high
diffusion rates of spin packets observed in Ref.~\cite{Aws_nature}. 
The effect of the metal-insulator transition on spin diffusion,
which can be substantial in semiconductors,\cite{MIT} is judged
in this circumstance to be small. 

Ineffective screening in semiconductors requires that local
variations 
in the conduction electron density ($\Delta n({\bf x})$) be,
under normal circumstances, balanced by
a local change in the valence hole density ($\Delta
p({\bf x})$). 
Even small local variations of charge in a
semiconductor produce large
space-charge fields which force the system to approximate local neutrality.
In metals, by contrast, local charge
density variations are screened out on length
scales of Angstroms. 
The $\Delta n({\bf x}) \sim \Delta p({\bf x})$
constraint in semiconductors has key implications for the motion of
packets of increased carrier density.\cite{Roosbroeck,Smith} 
If such a packet moves, both the conduction electrons and
valence holes which comprise it must move together.  The motion of holes in
semiconductors tends to be much slower (due to their lower mobility)
than that of electrons, so hole mobility and diffusion
tends to dominate the properties of a
packet consisting of both electron and hole density
variations. 

Spin packets in semiconductors are also subject to these constraints.
Consider a spin packet which involves an increase in the density of spin-up
electrons, or $\Delta n_\uparrow({\bf x}) > 0$.
In undoped semiconductors it is not possible for the population
of the other spin species to be substantially decreased, for the 
thermally generated background of conduction electrons is quite small.
Hence an increase in the population of one spin species of carrier
implies an increase in the total population of that carrier, so
$\Delta n_\uparrow({\bf x})>0$ implies $\Delta n({\bf x})>0$. The 
increase in the 
total electron density then implies a local increase in the hole
density to maintain $\Delta n({\bf x}) \sim \Delta p({\bf x})$. Even 
if the holes in the packet are not
spin polarized themselves, their presence
affects the motion of the spin-polarized electrons.

In a doped semiconductor, however, 
there is a substantial background of conduction electrons, so
$\Delta n_\downarrow({\bf x})$ can be significantly less than zero. Thus one
can create a spin packet through a spin imbalance in the
conduction band
($\Delta n_\downarrow({\bf x}) = -\Delta n_\uparrow({\bf x})$),
without excess electrons or holes
($\Delta n({\bf x}) = 0 = \Delta p({\bf x})$).\cite{Aws_PRL} 
This spin packet {\it does not drag
a local inhomogeneous hole density with it,} and thus its mobility and
diffusion properties are very different
from those of a spin packet in the undoped semiconductor. 

The two situations are distinguished in Fig.~\ref{chsp} for a
nonmagnetic
electron-doped material. Figure~\ref{chsp}(a) shows an inhomogeneous
electron-hole density in the form of a spatially localized packet. This
disturbance, which we will refer to as a charge polarization
packet (or charge packet), could be created optically, in which case the
excitation process guarantees $\Delta n({\bf x}) = \Delta p({\bf x})$, or by
electrical injection, in which case space charge fields 
force $\Delta n({\bf x}) \sim \Delta p({\bf x})$. This type of disturbance is
thus fundamentally multiple-band. Figure~\ref{chsp}(b), however, shows a
spin disturbance within the conduction band, which is
an enhancement of the density of spin-up electrons and a
corresponding reduction of the density of
spin-down electrons ($\Delta n_\downarrow({\bf x}) = - \Delta
n_\uparrow({\bf x})$). There is no corresponding inhomogeneity
in the hole density ($\Delta p_\downarrow({\bf x}) = \Delta
p_\uparrow({\bf x}) = 0$), so the disturbance in essence only
involves a single band.  This type of disturbance will be
referred to as a spin polarization packet, or a spin
packet. 

As described and demonstrated in Ref.~\cite{Aws_PRL}, generation of this 
spin packet can be performed optically with circularly polarized
light in a system where the spin relaxation time for holes
is short, and for electrons is long, relative to the
recombination time. Shortly after the excitation process creates spin-polarized
electrons and holes, the holes lose their spin polarization. 
During the recombination process the unpolarized holes annihilate
an equal number of spin up and spin down electrons, leaving behind
excess spin polarization in the conduction band.

We now describe the implications for mobility and diffusion
of these two types of packets.
The motion of a charge packet (Fig.~\ref{chsp}(a)) involves dragging
both a conduction and valence disturbance, and is described
by an ambipolar mobility and diffusion constant,
\begin{eqnarray}
\mu_a &=& {(n-p)\mu_e\mu_h\over n\mu_e + p\mu_h},\\
D_a &=& {n\mu_e D_h+p\mu_h D_e\over n\mu_e + p\mu_h},
\end{eqnarray}
where $D_e$, $\mu_e$ and $D_h$, $\mu_h$ are the
diffusion constants and mobilities for electrons
and holes respectively.
For $n$-doping ($n\gg p$),
$D_a\sim D_h$ and $\mu_a\sim \mu_h$, so diffusion and
mobility of the charge packet is
dominated by the {\it holes}. 
The mobility and diffusion constants of the spin packet of Fig.~\ref{chsp}(b),
however, are
\begin{eqnarray}
\mu_s &=& {(n_\downarrow+n_\uparrow)\mu_{e\downarrow}\mu_{e\uparrow} 
\over n_\downarrow\mu_{e\downarrow} + n_\uparrow\mu_{e\uparrow}},\\
D_s &=& {n_\downarrow\mu_{e\downarrow}D_{e\uparrow}+
n_\uparrow\mu_{e\uparrow}D_{e\downarrow}
\over n_\downarrow\mu_{e\downarrow} +
n_\uparrow\mu_{e\uparrow}},
\end{eqnarray}
where we now allow the different spin directions to have
different mobilities and diffusion constants. 

For the
nonmagnetic semiconductor of Ref.~\cite{Aws_nature},
with $n_\uparrow =
n_\downarrow$, $\mu_{e\uparrow}=\mu_{e\downarrow}$,
and $D_{e\uparrow} = D_{e\downarrow}$, the mobility and
diffusion constants of the spin packet are
merely the {\it electron} mobility $\mu_e$ and
diffusion constant $D_e$.  Thus the mobility of the spin
packet is predicted to be the same as that
measured in transport. 
The importance of the metal-insulator transition can be estimated
by considering $\sigma(L)$, the dependence of conductivity on the physical
length scale probed.\cite{MIT}
In Ref.~\cite{Aws_nature}
the mobility measured optically over a distance of microns was seen to be comparable
to the mobility from transport measurements through
the entire sample, suggesting that the material was not
sufficiently close to the metal-insulator transition to exhibit
significant effects on the conductivity on these length scales.

Because the diffusion and mobility of spin and charge packets
in doped semiconductors are determined by the properties of
a single carrier species, we can relate the
mobility $\mu$ of a packet 
to the diffusion constant $D$ describing
the spread of the packet with an expression\cite{Smith} derived for
a single species,
\begin{equation}
qD = -\mu{\int_0^\infty N(E)f_o(E)dE\over \int_0^\infty
N(E)\left(\partial f_o(E)/\partial E\right)dE}.
\end{equation}
Here $N(E)$ is the density of states of the band with the
zero of energy chosen so that the band edge is $E=0$, $f_o(E)$
is the Fermi function, and $q$ is the charge of the species.
In the low density limit $(\partial f_o(E)/\partial E) =
f_o(E)/kT$, where $T$ is the temperature and $k$ is
Boltzmann's constant, and so $eD=\mu kT$, which is
Einstein's relation. 

Figure~\ref{chspvdens} shows $eD/kT\mu$ for a spin
packet (solid line) and a charge packet (dashed line) in
$n$-doped
GaAs at $T=1.6$K. The relevant mobility and diffusion
constant for the spin packet are those of the conduction electrons,
while those for the charge packet are those of the valence
holes. 
This enhancement over the Einstein relation is directly
attributable to Fermi pressure, that is, the faster increase of the 
chemical potential with density for a degenerate Fermi gas relative
to a non-degenerate Fermi gas. For a given gradient in the density of
the degenerate Fermi gas, a larger gradient in the
chemical potential results, yielding faster diffusion.

Fermi pressure is
substantially more important for spin packets than for
charge packets, which exhibit the effect at densities closer
to 10$^{18}$~cm$^{-3}$ at low temperature, and may require
densities as high as 10$^{20}$~cm$^{-3}$ at room
temperature.\cite{VanDriel} 
At higher temperatures it also requires
considerably higher densities for Fermi pressure to play a
significant role in spin packet diffusion, but the densities
are still achievable, corresponding to 10$^{18}$~cm$^{-3}$.
The quantitative difference in the significance of Fermi pressure
for spin packets, which are dominated by 
conduction electron properties, and for charge packets, which are dominated
by valence hole properties, occurs
because the conduction band has a 
density of states $(m_e/m_h)^{3/2}\sim 0.045$ smaller than the valence
band in GaAs and therefore becomes degenerate at
lower density. At 10$^{16}$~cm$^{-3}$ the enhancement over
the Einstein relation is 12, which is in good agreement with the
``more than one order of magnitude'' enhancement seen in
Ref.~\cite{Aws_nature}. 

In order to generate spin packets in a $p$-doped semiconductor the
time scales of spin decoherence in the conduction and valence
band would need to be different, but perhaps a semiconductor will be
found where this is possible. In this case it is the charge packet
which is dominated by the diffusion and mobility properties
of the conduction electrons, whereas the spin packet is
dominated by the properties of the valence holes. Thus the charge
packet is over an order of magnitude more mobile than the
spin packet, precisely the opposite case as for an $n$-doped semiconductor.

\begin{figure}
\caption[]{Ratio of diffusion to mobility for spin packets
polarized parallel (dashed) and antiparallel (solid) to the equilibrium
carrier spin polarization of the semiconductor Be$_x$Mn$_y$Zn$_{1-x-y}$Se at
1.6K and 30K as a function of conduction electron density 
(Be doping). The conduction mass is $0.16m_o$ and the heavy
hole mass is $0.74m_o$, where $m_o$ is the free 
mass.\cite{BeMnZnSe}\label{spsavdens}}
\end{figure}
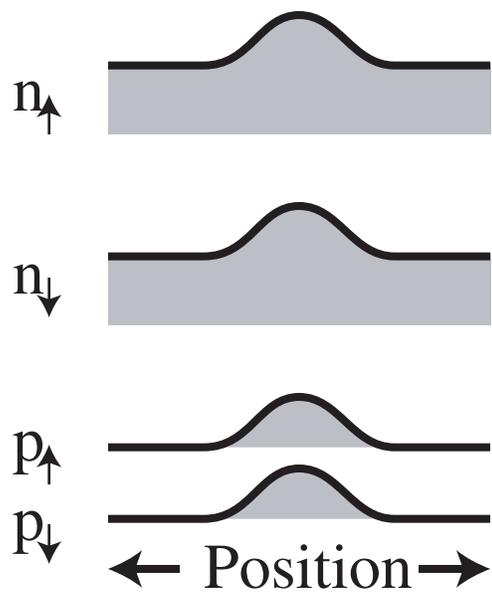
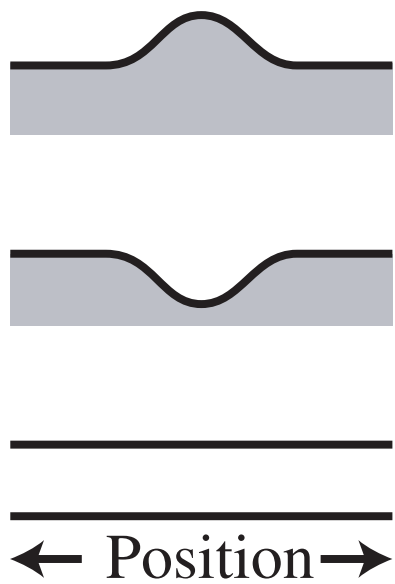
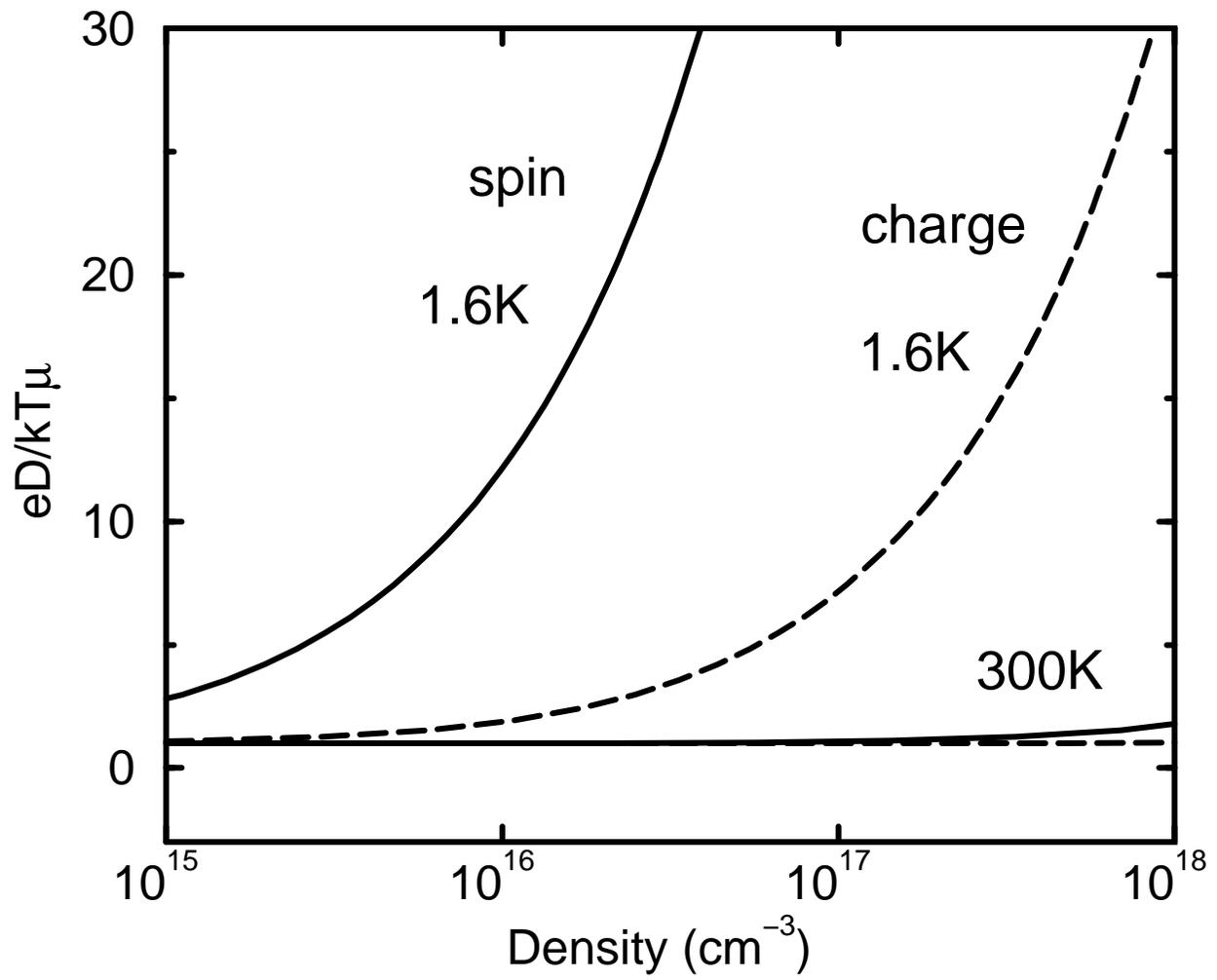
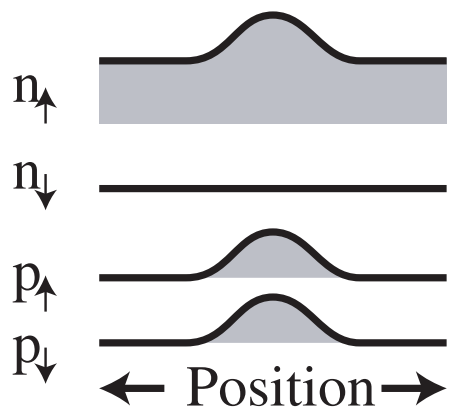
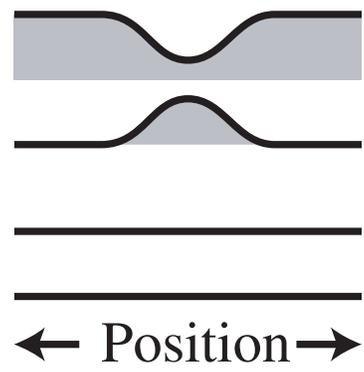
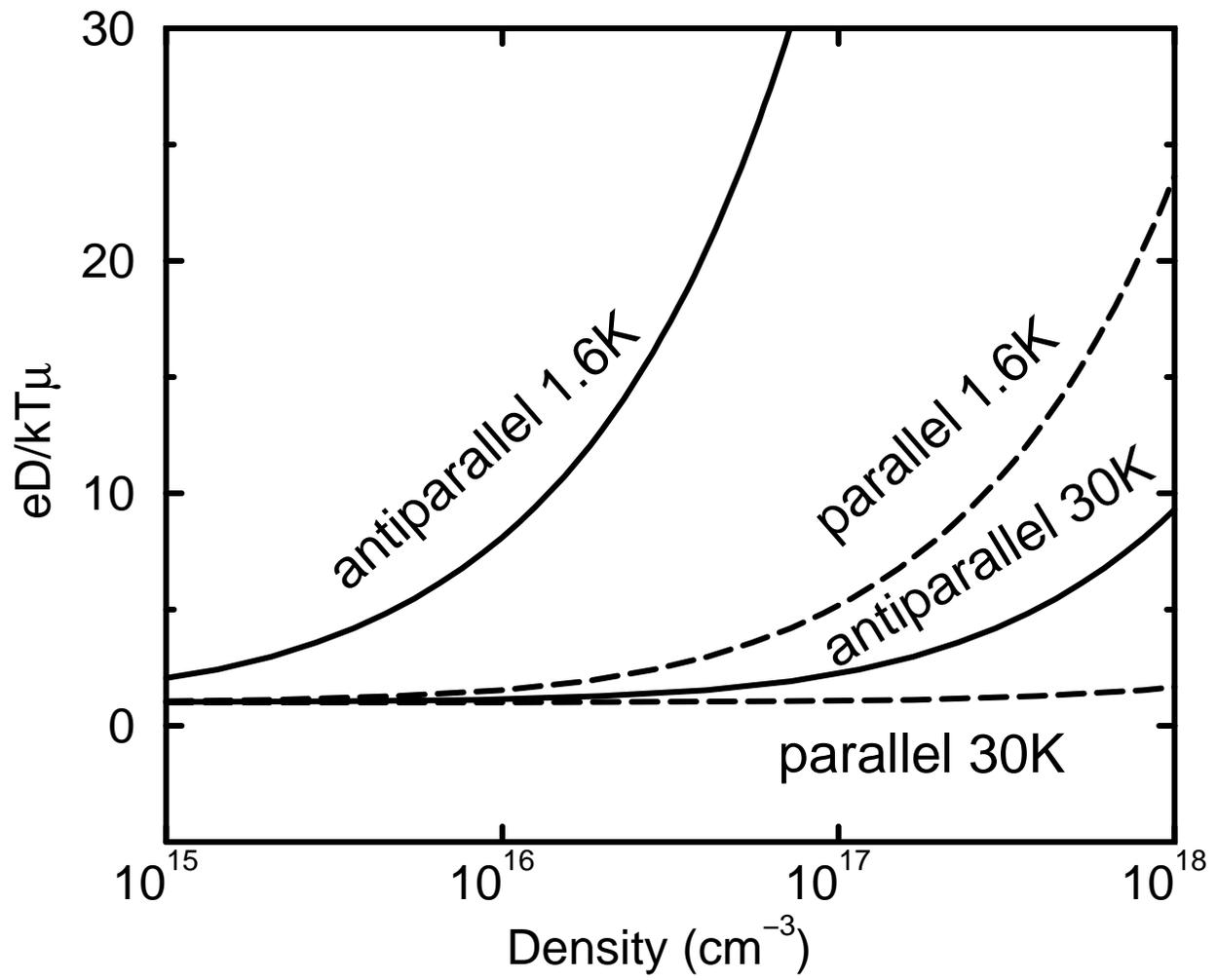

We now turn to the behavior of spin and charge packets in a
spin-polarized semiconductor, where equilibrium densities,
mobilities and diffusion constants can differ for the two
spin densities. Our first specific
example will be an $n$-doped spin-polarized semiconductor (such as
BeMnZnSe)
which we assume
is 100\% spin polarized at the chemical potential. We note that the 
spin splitting required to achieve this polarization in a semiconductor,
where typical Fermi energies are $10-100$~meV, is much less
than that required in a metallic system.
For this semiconductor
in equilibrium
$n_\uparrow>0$, but $n_\downarrow$, $p_\uparrow$, and
$p_\downarrow$ are all approximately zero. 
As shown in Fig.~\ref{spsa} a single-band spin polarization
packet
is only possible for a spin packet polarized antiparallel to the
equilibrium carrier spin polarization.
This restriction occurs because
$\Delta n_\uparrow({\bf x})<0$ is possible, but 
not $\Delta n_\downarrow({\bf x})<0$. Thus a packet with
spin polarized parallel to the equilibrium spin (Fig.~\ref{spsa}(a))
must consist
of both electron and hole perturbations ($\Delta n_\uparrow({\bf x}) >0$
and $\Delta p({\bf x})>0$) and would have
diffusion and mobility properties dominated by the
minority holes. The antiparallel spin packet (Fig.~\ref{spsa}(b)), 
however, can be a single
band disturbance with $\Delta n_\uparrow({\bf x})<0$ and 
$\Delta n_\downarrow({\bf x})>0$. Such a spin packet would have
diffusion and mobility properties entirely determined by
those of the majority electrons, and thus over an order of
magnitude faster. We show in Fig.~\ref{spsavdens} the
different ratios of diffusion constant to mobility for spin
packets polarized parallel and antiparallel to the equilibrium
carrier spin polarization.

The behavior of spin packets in a $p$-doped spin-polarized semiconductor,
such as \hbox{GaMnAs,} is completely the opposite.
Here a spin packet polarized parallel to the equilibrium 
carrier spin polarization
would require a conduction electron component.
The minority
carriers (the electrons) would determine the mobility and
diffusion constant of such a packet.
A spin packet polarized antiparallel to the equilibrium
carrier spin polarization
could consist entirely of holes, however,
and would have a much smaller mobility and diffusion constant. 
This qualitative difference in the
diffusion and mobility of spin polarization packets in the
$n$ and $p$-doped 
semiconducting systems should have technological
implications for spin electronic devices.

We conclude with a brief comment on the behavior of spin distributions
in inhomogeneous semiconductors compared to those in metallic ferromagnets.
As pointed out in Ref.\cite{Hershfeld}, in metallic ferromagnets the
short-distance physics of screening can be entirely separated from the
physics of spin populations by writing a drift-diffusion equation for the
chemical potential rather than the density. This separation depends on
the linear dependence of the density on the chemical potential in these
systems. This relationship does not hold in semiconductors and 
the separation of the screening length scale from the spin
distribution length scale is no longer possible. Thus the exploration of 
spin transport in inhomogeneously doped spin-polarized semiconductor
materials should yield a rich range of behavior distinct from metallic
systems.

One of us (M.E.F.) would like to acknowledge the support of the Office of 
Naval Research through Grant No. N00014-99-1-0379.

\begin{figure}
\caption[]{Spin subband profile of charge and spin polarization packets in a 
semiconductor. Space charge constraints require $\Delta n({\bf x})\sim
\Delta p({\bf x})$, so a charge disturbance in the conduction band 
requires a corresponding disturbance in the valence band. In contrast,
a pure spin disturbance can exist in the conduction band of a doped
semiconductor without any disturbance in the valence band.\label{chsp}}
\end{figure}

\begin{figure}
\caption{Ratio of diffusion to mobility for charge (dashed)
and spin packets (solid) in GaAs
at 1.6K and 300K as a function of background conduction electron ($n$)
density. The diffusion and mobility of the
charge packet is dominated by the valence holes, whereas that of the 
spin packet is dominated by the conduction electrons.\label{chspvdens}}
\end{figure}

\begin{figure}
\caption[]{Spin subband profile of spin polarization packets
polarized parallel and antiparallel to the equilibrium
carrier spin polarization 
of an $n$-doped spin-polarized semiconductor. \label{spsa}}
\end{figure}

\end{document}